*Switching on and off the Risken-Nummedal-Graham-Haken instability in Quantum Cascade Lasers*

A.V. Antonov,[1] D.I. Kuritsyn,[1] A. Gajic,[2] E.E. Orlova,[1] J. Radovanovic,[3] V.V. Vaks,[1] and D. L. Boiko[4]

[1]Institute for Physics of microstructures, Nizhnij Novgorod, Russia

[2]Regulatory Agency for Electronic Communications and Postal Services, Palmotićeva 2, 11000 Belgrad, Serbija

[3]School of Electrical Engineering, University of Belgrade, Bulevar kralja Aleksandra 73, 11120 Belgrade, Serbia

[4]Centre Suisse d'Electronique et de Microtechnique SA (CSEM), CH-2002 Neuchâtel, Switzerland

**Abstract:**

We report on experimental observation of low-threshold Risken-Nummedal-Graham-Haken (RNGH) instability in mid-infrared InGaAs/InAlAs Quantum Cascade Lasers (QCLs) operating at 8 µm wavelength. The devices employ sequential resonant tunneling and can provide optical gain on the diagonal or vertical transition. Depending on the tunneling resonance detuning, the lasers can operate either on a stable part of I-V curve or with unstable electric field domain formation. We report how all these features impact the occurrence of the RNGH instability with broadband multimode emission, and show how they can be tailored with an applied bias or via optical excitation of free carriers. We present the results of 2$^{nd}$ order interferometric autocorrelation measurements and time-resolved spectral measurements in free-running QCLs and QCLs subjected to optical pumping. We explain the diversity of dynamic regimes in QCLs using a simple analytical expression for the second threshold.

Key words: RNGH instability, QCL MIR laser

# I. INTRODUCTION

Mid infrared (MIR) spectral range is attractive for various applications in security, health, gas sensing and gas analysis [1]. Single mode and narrow spectrum Quantum Cascade Lasers (QCLs) provide access to spectral measurements of molecular fingerprints (ro-vibrational bands) in the MIR range. However Fabry-Perot (FP) cavity QCLs often exhibit a broadband emission with a large number of excited cavity modes, which is associated with low-threshold multimode Risken-Nummedal-Graham-Haken (RNGH) instability [2,3]. Such frequency combs being additionally stabilized via driving current modulation and by the design of QCLs with reduced group delay dispersion find market applications in MIR spectroscopy [4]. Nevertheless none of these combs in stand-alone single-section QCLs have been converted in a pulse train in the time domain. (In contrast to this, an external cavity should be used in order to provide a "memory effect" in the time domain and to overcome the problem of very fast gain recovery in QCL [33,34]). This indicates that phases of different modes in the comb are not all locked to each other as in a classical mode-locking regime. Therefore strong spectral broadening and multimode emission in QCLs occurs due to another mechanism, such as parametric RNGH gain instability. However the cause of multimode RNGH instability in QCLs so far remains unclear and was a highly debated subject [1,2,5,6,7]. As a consequence, little was certain about practical approaches to tailor the multimode behavior of QCL devices for specific application needs.

In this paper we report on experimental observation of excitation and suppression of low-threshold RNGH instability in mid-infrared InGaAs/InAlAs QCLs with level alignment and resonant tunneling between injector and active quantum wells (QWs). Performing 2-d order interferometric autocorrelation measurements and time-resolved spectral measurements in free-running QCLs and QCLs subjected to optical pumping we have identified two different mechanisms that can be used for tailoring RNGH instability with applied bias field and/or optical excitation of free carriers. The first mechanism is due to switching between a diagonal transition from injector to active region and a vertical transition within active QWs during a transient turning-on process. The second mechanism is based on unstable electric field domain (EFD) formation [30,31] when the applied bias field is greater than that at the tunneling resonance.

Our experimental results are in agreement with the analysis [6,7] that attribute the low-threshold RNGH instability in QCLs to the gratings of population inversion and coherences (medium polarization) induced by the standing cavity mode pattern. More specifically, the analytic expression for the threshold of instability (second threshold) from [6] involves the population difference lifetime $T_1$ (or gain recovery time), the dephasing time $T_2$ and the carrier diffusion coefficient $D$ that impact formation of the carrier and coherence gratings. Within the framework of this model, we attribute the onset of the RNGH instability at the turning-on process to the difference in the gain relaxation times $T_1$ on the diagonal (injector-active QWs) and vertical (within active QWs) transitions. On the other hand, the suppression of RNGH instability under EFD formation is explained within the framework of our model by the washing out of the induced gratings by the inhomogeneous bias field and carrier distributions.

The paper is organized as follows. In Section II we provide a brief review of the QCL second threshold analysis and available experimental reports on RNGH instability in QCLs. In section III we describe the samples used in this study. Section IV reports on the emission dynamics in free-running QCLs and QCLs subjected to additional interband optical pumping, observed both in the steady-state operation and in the transient regime after turning-on the pump current. The last Section V concludes the paper with a brief wrap-up summary.

**II. RNGH instability in QCLs**

**A. Theoretical background**

The multimode RNGH instability [9,10] in a continuous traveling wave single-mode laser (e.g. unidirectional ring laser) arises due to parametrically induced Rabi splitting of the saturated modal gain curve [11]. As a result of such spectral reshaping, the laser can provide sufficient optical gain for small perturbations in other longitudinal modes. In general, the multimode RNGH instability is related to excitation of rapid coupled oscillations of the medium polarization and population inversion. They show up as quasi-periodic (chaotic) or regular self-pulsations in the output laser emission, whereas the optical spectrum is split in mode clusters (sidebands) or reveals strong broadening on the order of the Rabi oscillation frequency. While the second threshold in unidirectional ring lasers is expected to be at the pump rate 9 times higher than that at the lasing threshold [9,10], the FP cavity QCLs reveal second threshold at a current of only 5 – 10 % higher than the lasing threshold [2,3,5].

The main difference between the two classes of lasers is due to the standing mode pattern in FP cavity lasers yielding a spatial hole burning (SHB) effect [see below a discussion to Eq.(2)]. Thus in Refs.[6,7] the second threshold reduction in QCLs is attributed to the mode coupling through the induced gratings of population inversion and coherences (medium polarization). The following expression for the second threshold was obtained for the long-cavity QCLs [6]:

$$p_{th2} = 1 + \frac{T_{2,eff}}{T_{2,g}} + \left( \frac{T_1}{T_g} \frac{T_{2,eff}}{T_{2,g}} - \frac{T_g^2}{T_{2,g}^2} \right) \left[ \sqrt{1 + 2\frac{T_{2,g} T_{2,eff}}{T_g^2}} - 1 \right], \quad (1)$$

where $p$ is the pump current normalized to the lasing threshold. The overall relaxation times for the carrier population grating $T_g = \left(T_1^{-1} + 4Dk^2\right)^{-1}$ and the coherences grating $T_{2,g} = \left(T_2^{-1} + 9Dk^2\right)^{-1}$ as well as the effective transverse relaxation time $T_{2,eff} = \left(T_2^{-1} + Dk^2\right)^{-1}$ account for the effect of carrier diffusion; $D$ is the diffusion coefficient for electrons in the plane of active QWs and $k=2\pi n_g/\lambda$ is the wavenumber of the main lasing mode in the cavity. Equation (1) is valid when the cavity free spectral range (FSR) is much smaller than $T_g^{-1}$ (e.g. in mm-length QCLs). For the case when $T_{2,g} T_{2,eff} \ll T_g^2$ the following approximation can be used:

$$p_{th2} \approx 1 + \frac{T_{2,eff}^2}{T_g^2} \left( \frac{1}{2} + \frac{T_1}{T_g} \right). \quad (2)$$

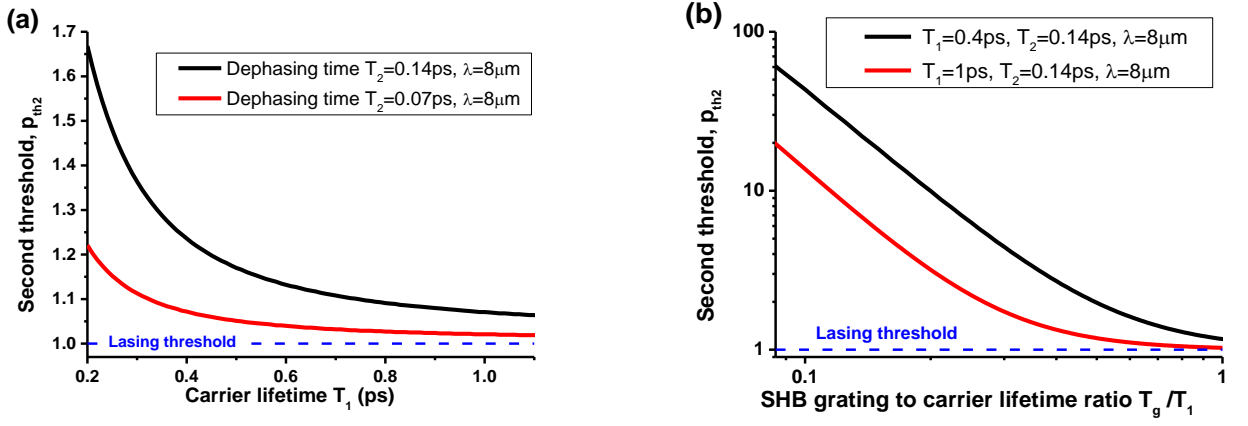

**Fig. 1.** (a): Normalized second threshold vs carrier relaxation time $T_1$ calculated for the carrier decoherence time $T_2$ of 70 fs (red curve) and 140 fs (black curve), diffusion coefficient $D=180$ cm$^2$/s. (b): Normalized second threshold vs SHB grating relaxation time for population difference lifetime $T_1$ of 0.4 ps (black curve) and 1 ps (red curve), decoherence time $T_2=140$ fs. The horizontal dashed line indicates the lasing threshold. The emission wavelength $\lambda=8$ µm.

The last expression allows us to elucidate the role of the SHB effect. For instance in MIR QCLs, the diffusion length of the electrons in the plane of active QWs is of $\sqrt{DT_1}\sim 0.1$ µm, which is much smaller than the wavelength. Therefore the induced population grating has a large contrast and lowers the 2$^{nd}$ threshold. Indeed, if we assume that $Dk^2 \ll T_2^{-1}, T_1^{-1}$ in Eq.(2) and thus neglect the diffusion, we find that the seconds threshold in QCLs is just above the lasing threshold $p_{th2}^{(QCL)} \to 1 + 3T_2^2/2T_1^2$ provided $T_2/T_1 \ll 1$.

In order to illustrate the role of the SHB effect, QCLs should be compared to the interband QW (or bulk) laser diodes (LDs) operating in the visible (VIS) and near-infrared (NIR) spectral ranges. For LDs with the characteristic gain recovery time of $T_1\sim 1$ ns and bimolecular diffusion coefficient of $D\sim 10$ cm$^2$/s, an estimation of the typical carrier diffusion length yields the value of $\sqrt{DT_1}\sim 1$ µm. The diffusion length exceeds the wavelength in the gain medium and hence the induced population grating has a low contrast. Assuming that $Dk^2 \gg T_2^{-1}, T_1^{-1}$ in (2), we find that $p_{th2}^{(LD)} \to 9 + 16T_1/T_g$. We thus partially recover the conclusion of Risken, Nummedal, Graham and Haken that the multimode instability requires at least 9-fold pump excess above the lasing threshold. However in the case of LDs, the second threshold is much higher because $T_1/T_g \gg 1$. Thus as a result of the shorter emission wavelengths and longer (bimolecular) diffusion lengths of the carriers, FP cavity QW LDs do not show RNGH instability at experimentally achievable pumping currents [23]. [Obviously, these considerations do not apply to the case of quantum dot (QD) and quantum dash (QDash) laser diodes where the diffusion relaxation of the gratings along the cavity waveguide is prevented by the quantum confinement and the characteristic gain recovery time is on ps scale.]

Following along these lines, the mechanism of the RNGH instability switching due to a change of the gain recovery time is illustrated by Figure 1(a), where the behavior of the second threshold (1) in the range of typical carrier lifetimes for MIR QCLs is plotted. The diffusion coefficient used in this figure allows the formation of the population and coherence gratings and corresponds to QCL operating with equal voltage drops on all quantum cascade periods, within a stable part of I-V curve with positive differential resistance [30]. In these conditions, a sudden increase of the carrier lifetime $T_1$ (e.g. as a result of the lasing transition change) causes a drop of the second threshold, so that the QCL initially operating in the usual regime with emission of a few cavity modes will switch to broadband multimode RNGH self-pulsations.

The second mechanism of the RNGH instability switching due to the change of the contrast of the induced population grating is illustrated in Figure 1(b). Vanishing of the population grating can occur in QCL operating at a bias higher than at a tunneling resonance, which would correspond to the unstable part of I-V curve with negative differential resistance if the voltage drops on all periods would be the same. However instead of this, when a QCL is driven by a current source, the unstable part of I-V curve leads to a formation of electric field domains (EFDs) with non-uniform charge accumulation and depletion across multiple QCL periods [30,31]. We assume that these inhomogeneities in the carrier distribution reduce the efficiency of the mode coupling through the induced carrier grating, the situation which is simulated in Fig. 1(b) by allowing the grating relaxation time $T_g$ to decrease well below $T_1$.

The contrast reduction of the induced population grating leads to the increase of the second threshold to prohibitively high currents.

In the present paper we report on the experimental studies supporting both proposed mechanisms of RNGH instability switching in QCLs.

**B. Experimental studies of RNGH self-pulsations in QCLs**

Strong spectral broadening in single-section FP cavity QCLs was reported in Refs.[2,3,5,14]. It was observed in the ridge waveguide and in the buried heterostructure devices operating under similar driving conditions. In Refs. [2,3,14] the spectral broadening was associated with RNGH multimode instability and in Ref. [2,14] it was attributed to a combined effect of spatial hole burning and saturable absorption features present in QCL cavity. However, the nature of such saturable absorption which should be equally strong in a ridge waveguide and a buried heterostructure QCLs has never been fully clarified [1].

Another mechanism of instability in single-section QCLs have been proposed in Ref. [5], where it was linked to a parametric four-wave mixing gain instability, in full analogy with the phase-locked comb production in high-finesse optically pumped micro-cavities. Interestingly, this mechanism should lead to a soliton formation in the time domain [15]. Unfortunately very few measurements of the second-order interferometric autocorrelation (IAC) traces have been reported for QCLs [2,14]. They don't provide an unambiguous conclusion about the output QCL waveform and eventually about the pulse shape. None of the previous reports indicated IAC variation in function of the pump current which otherwise would greatly streamline the identification of the dynamic behavior of QCLs in the time domain.

All second-order IAC traces in DC-driven single-section FP cavity QCLs reported so far show the peak to background ratio close to 8:3, attesting rather a noisy (multimode) continuous wave (CW) lasing behavior [13] as opposed to formation of a pure phase-locked comb. Furthermore, none of the reported RF power spectra of intensity have shown a comb with multiple harmonics of the fundamental cavity frequency which would confirm the phase locking of the optical modes [24,25]. In contrast to this, all RF power spectrum reports are limited to the beat note at fundamental frequency.

The experimental data reported in the literature show that RNGH instability in QCLs does not always lead to Rabi splitting of the optical mode spectrum and just broadens the lasing spectrum up to the width of the Rabi oscillations sidebands. The QCL spectral behavior depends on the operation temperature [2,3,14]. For example, the lasing spectra in a buried heterostructure QCL [2] measured at low temperatures (80-150 K) are almost uniform, whereas at higher temperatures the Rabi splitting between the mode clusters becomes apparent. At room temperature two distinct mode clusters emerge in the optical spectrum, while the overall spectral width does not change, remaining of ~60 cm$^{-1}$ at two-fold excess of the pump current above the lasing threshold ($p$~2). The thermal population of injector levels yielding a temperature-dependent saturable absorption has been suggested as a possible mechanism of such spectral reshaping [2]. However, a very different dependence of the spectral behavior of QCLs on the operation temperature have been observed in [3]. The optical spectra in [3] reveal the onset of the lasing mode clustering at a liquid He temperature (6 K), while at an increased temperature of 77 K, no spectral reshaping is observed in the broad multimode emission, which remains of ~35-40 cm$^{-1}$ overall width for the current of 1.7-1.8 times above the lasing threshold. Although available experimental reports do not elucidate the nature of the temperature effect, they allow us to spot one common feature: once a QCL sample exhibits clear Rabi splitting at some temperature, it also reveals a broad multimode emission (either with or without mode clustering) at other temperatures. The overall spectral width practically does not change with the temperature. We believe that the formation of a broad multimode emission (or a comb) with the overall spectral width of ~2$\Omega_{Rabi}$ as opposed to the appearance of the mode clusters on the Rabi-frequency sidebands is governed by the group delay dispersion in the cavity [4]. Most importantly, in both cases the origin of the multimode emission is related to RNGH instability. For spectroscopic applications of QCL combs, serious design efforts have been made to optimize active QWs, reduce group velocity dispersion and enlarge the spectral width of the gain curve on the active vertical transitions by incorporating several QWs with different transition energies [4].

In the present work we study the effect of switching between the diagonal transition injector-active region and the vertical transition in the active QWs as well as the effect of unstable EFDs formation [30,31] due to misalignment of injector ground state and the upper laser level. The switching–off behavior of the multimode RNGH instability at high bias currents was observed in [3], where it was attributed to a decrease of the optical power in the cavity due to a rollover of the QCL output characteristic [29]. We argue that a gradual decrease of the output power would rather produce a gradual decrease of the spectral width prior to a collapse of the multimode RNGH instability. However the experimentally observed cessation of the broadband RNGH self-pulsations was very abrupt.

## III. QCL SAMPLES WITH SEQUENTIAL RESONANCE TUNNELING

As a model system for our studies on RNGH instabilities in QCLs we choose the InGaAs/InAlAs on InP design with layer composition and doping profile of the sample N655 from Ref. [8]. Its electronic band structure is depicted in Fig. 2(a). The active region of one fundamental period consists of four quantum wells (QWs) followed by a collector / injector of the next period. The periods are separated by injection barriers and are pumped via sequential resonant tunneling [8,20,21,29]. The structure [8] utilizes strong coupling between the injector subbands and the active levels in the QWs and is designed to operate at the vertical transition 3-2 in the active QWs. The upper $/3>$ and lower $/2>$ lasing transition subbands, the two lowest-energy injector subbands (the ground state $/i>$ and the first excited state $/i´>$) and the two highest-energy collector subbands (the top excited state $/c>$ and the first de-excited state $/c´>$) are of particular interest. The transition energies between these states are depicted in Fig. 2(b) in function of the applied bias. They are calculated by solving the Schröedinger equation for an injector-active region-injector domain using the finite difference method (FDM) [35] and taking into account the material dependent effective masses and the energy band nonparabolicity. The periodic voltage drop (PVD) conditions with equal voltage drops on all periods are assumed. This approximation is valid on the positive differential resistance (PDR) part of the I-V curve. The model predictions for the negative differential resistance (NDR) part shall be used with caution, admitting the effect of electric field domains (EFDs) that renders field and carrier populations inhomogeneous across the periods [30,31].

The two lowest-energy diagonal transitions ($i$-2 from the lowest-energy injector subband and 3-$c$ to the highest-energy subband in the collector) are red shifted from the main vertical transition. However the two next diagonal transitions ($i'$-2 from the injector first excited state and 3-$c'$ to the first de-excited state in collector) exhibit a blue shift from the main lasing transition. This blue shift provides an important hint for interpretation of experimentally measured spectra.

The carriers from the lower lasing level in the active region relax via intersubband double LO phonon scattering. We estimate the carrier transport characteristics with a scattering intersubband model [26], which takes into account both electron-LO phonon and electron–electron scattering, interface roughness, assuming equal voltage drops on all of the periods and charge neutrality. The model provides the subband populations and the intersubband scattering rates by self-consistently solving a system of rate equations. Figure 2(c) shows the total current as a function of applied bias field (blue curve, left axis). Although our transport model is simplified [8,16,22], it provides sensible results at room temperatures, when the transport is dominated by incoherent scattering [20].

In agreement with the model from Ref. [8], our simplified model confirms the tunneling resonance at the bias field of 44 kV/cm [see the second local maximum on the total current curve in Fig. 2(c)] due to the alignment between the first excited state $|i'\rangle$ in the injector and one of the upper levels in the active QWs. Due to their complex band structure with multiple subbands, superlattices and quantum cascade structures usually exhibit multiple tunneling resonances [27]. We find another resonance at ~34 kV/cm bias due to alignment between the ground injector level /*i*> and the upper lasing level /3> [the first local maxima on the blue curve seen in Fig. 2(c)]. But most surprisingly, the matrix element $z_{i2}$ of the diagonal transition *i*-2 at this resonance is only by a factor of 2 smaller than $z_{32}$ for the vertical transition. Since the population of the ground injector subband is higher, both transitions can provide comparable (unsaturated) optical gain (Fig. 2(c), right axis). Similar considerations apply to the optical gain on the transitions *i´*-2 (from the first excited injector subband) and on the transition 3-*c´* (to the first de-excited state in the collector). Their small-signal gain curves are also plotted in Fig. 2(c) (right axis).

In Figure 2(c), the lasing effect on the vertical transition 3-2 (black curve) is possible starting from the bias field of ~32 kV/cm (when the total current reaches the lasing threshold of ~ 1.6 kA/cm$^2$) and continuing at higher bias values (more details are given in Section IV.C). The emission on the diagonal transition *i*-2 may occur well above the lasing threshold, in the bias range from ~33 to ~34.5 kV/cm (brown curve). On the other hand, the diagonal transitions *i´*-2 and 3-*c´* provide an important optical gain near the lasing threshold, at the bias around 32 kV/cm (red and green curves). Since the energy splitting between these transitions and the vertical transition 3-2 is much smaller than the thermal energy *kT* (see Fig. 2(b)), these results indicate a possibility of the lasing transition switching. Note, however, that our model does not take into account the photon induced carrier transport [16]. Moreover, we have assumed a stationary regime, while some of the experimental observations are made for a transient behavior when the carrier density is larger than at equilibrium and hence the gain coefficients are higher.

In experiment, the lasing transition can be determined from the emission wavelength and its sensitivity to the bias field. Diagonal transitions usually exhibit a linear quantum confined Stark effect (QCSE) energy shift [28], while the vertical transitions are sensitive to electric field only in the second order of the perturbation theory. In Fig. 2(b), this simple picture is warped by the resonant interaction of the subbands. Nevertheless with the bias in the range of the PDR (32-34 kV/cm), all diagonal transitions in Fig. 2 (b) show stronger up-chirps than the vertical transition 3-2.

The upper levels of the considered transitions (the subbands $/i>$, $/i´>$ and $/3>$) have very different lifetimes. Respectively, the population differences on these transitions relax with dissimilar characteristic times $T_1$. In Fig. 2(d), their relaxation times $T_1$ are estimated along the lines of a conventional expression for the vertical transition 3-2 [1]:

$$T_1 = \tau_3 (1 - \tau_2/\tau_{32}), \tag{3}$$

where $\tau_2$ is the lifetime of the lower level, $\tau_3$ is the lifetime of the upper level and $1/\tau_{32}$ is the rate of direct scattering process from the upper to the lower level. For example, the diagonal transition $i´$-2 reveals a very short gain relaxation time $T_1$ at any bias [Fig. 2(d)], while the diagonal transition $i$-2 has a short relaxation time in the range of bias voltages corresponding to large optical gain [the range of 33-34 kV/cm in Fig. 2(c)]. In accordance with the data in Fig. 1 (a), these diagonal transitions exhibit much higher second thresholds compared to that of the vertical transition 3-2. Thus there is a range of bias fields where optical gain (and lasing) is possible both on the vertical and on the diagonal transitions, but RNGH instability may occur only for the vertical one. In Fig. 2(d) we do not yet take into account that the dephasing times $T_2$ for the two transitions are different, which would introduce further dissimilarities.

At the onset of NDR (at 34kV/cm), a single QCL period driven from a current source would exhibit a hysteresis loop. Thus for an increasing current, the operation point would jump to the next PDR part of the I-V curve (in our case it corresponds to the shoulder of the second tunneling resonance at 44 kV/cm bias). In a QCL with multiple cascades, the NDR leads to unstable EFD formation characterized by non-uniform charge accumulation and depletion across multiple periods [30,31]. A coarse approximation would be to assume that different QCL periods operate with the optical gain coefficients corresponding to different biases in Fig 2(c). However this picture does not account for non-uniform distribution of carriers. We can speculate that inherent inhomogeneities along the 3 mm cavity waveguide together with the charge accumulation layer due to EFDs, which is moving between the periods, wash out the induced spatial gratings. According to the simulations presented in Fig. 1(b), this will suppress RNGH self-pulsations by rising the second threshold to prohibitively high currents.

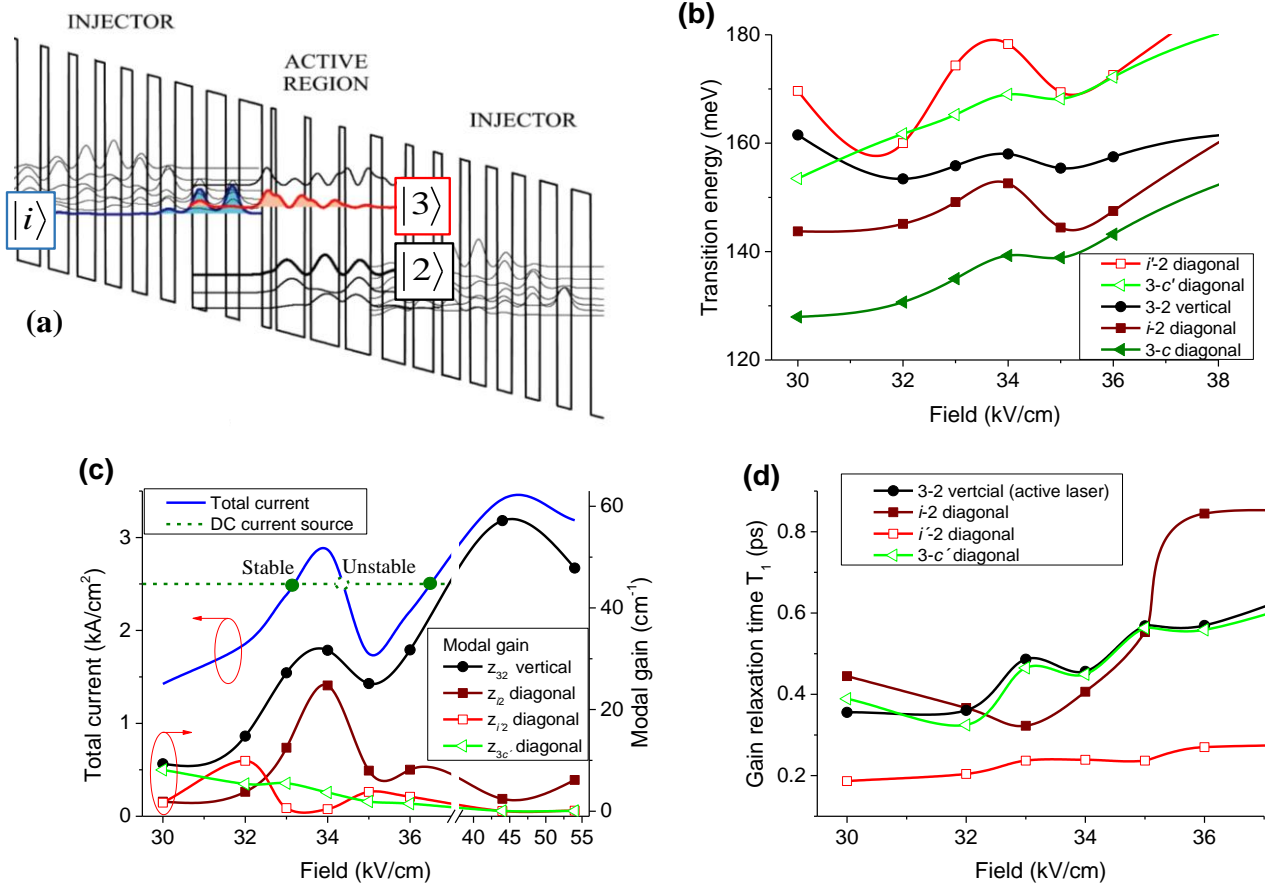

**Fig. 2.** (a) Band structure of one fundamental period of QCL for the bias field of 34 kV/cm calculated for the temperature 300 K. (b) Photon energies for vertical transition 3-2 in the active QWs (black solid curve) and for the four nearest diagonal transitions. (c) Total current density (blue curve, left axis), and the small-signal optical gain for the vertical 3-2 and a few diagonal transitions (solid curves, right axis) as a function of the bias field. Expected threshold conditions: 1.6 kA/cm2 at 32 kV/cm for the vertical transition in QCL of 3 mm long cavity (16 cm$^{-1}$ threshold gain) assuming $T_2$=0.13 ps and $D$=180 cm$^2$/s. (d) Gain recovery time $T_1$ in Eq.(3) for the vertical 3-2 and a few diagonal transitions.

In this study we use two nominally identical QCL samples (labeled as #1803 and #6326) with the epitaxial composition reproducing the four-QW design of the sample N655 from Ref. [8]. Both QCLs have a ridge waveguide of 3 mm long and 12 µm wide with HR coating on the back facet. The lasing wavelength is around 8 µm. The threshold current and the voltage drop are of ~0.5 A and 7.2 V respectively at -20°C heatsink temperature. Note that the threshold current density ~1.4 kA/cm$^2$ and the bias field ~30 kV/cm are in reasonable agreement with the model [see Fig. 10(b) below].

## IV. RESULTS AND DISCUSSIONS

### A. Steady-state free-running regimes

Figure 3 shows the variations of the lasing spectra with the DC pump current in the sample #1803 at two different temperatures. The laser was driven from a source measure unit (Keithley 2400) operated in the current source mode. The optical spectra are taken using Fourier Transform Infra-Red (FTIR) spectrometer (Vertex 80V, Bruker) equipped with a standard MCT detector (D316, Bruker) and operated in a rapid scan mode with the resolution of 0.1 cm$^{-1}$. As expected, the threshold current varies significantly with the temperature ($I_{th}$~0.5 A at -21°C and 0.6 A, that is 20 % higher, at -2°C). At both temperatures the laser shows broad multimode emission spectrum with two or more distinct mode clusters starting from the currents at ~10 % above the lasing threshold and up to 800-850 mA. The emission spectrum rapidly broadens with the current and reaches the width of 20 cm$^{-1}$ (600 GHz). In Fig. 3(a) we plot the expected modulation sidebands (dashed black curves) with the Rabi frequency estimated using the expression [6]:

$$\Omega_{Rabi} = \sqrt{\frac{p - v_0(p)}{T_1 T_{2\_eff}}}, \qquad (4)$$

where $p$ is the pump rate excess above the lasing threshold, and the parameter

$$v_0 = \frac{1}{2}\left(p + 1 + \frac{T_{2\_eff}}{T_{2\_g}} + \frac{2T_1 T_{2\_eff}}{T_g T_{2\_g}}\right) - \sqrt{\frac{1}{4}\left(p + 1 + \frac{T_{2\_eff}}{T_{2\_g}} + \frac{2T_1 T_{2\_eff}}{T_g T_{2\_g}}\right)^2 - p\left(1 + \frac{T_{2\_eff}}{T_{2\_g}}\right) - \frac{2T_1 T_{2\_eff}}{T_g T_{2\_g}}} \qquad (5)$$

accounts for the threshold current increase due to the spatial hole burning. Good agreement between the measured spectral widths and Rabi oscillations sidebands attests for the development of ultrafast non-adiabatic coherent dynamics in the laser and allows us to assume that experimentally observed broad emission spectra can be caused by RNGH instability. Below we provide further evidences to this, e.g. by measuring a change in the 2$^{nd}$ order interferometric autocorrelation functions. For the sake of brevity, from now on we will label this regime as RNGH self-pulsations.

In Fig. 3, the current is increased in steps of 50 mA. The multimode spectra collapse abruptly at the current values of 800-850 mA at both temperatures. The difference of the critical currents at the two temperatures is thus less than 50 mA. It is significantly smaller than the lasing threshold difference, which is strongly affected by the temperature. Therefore, the collapse of the broadband multimode emission cannot be regarded as a thermal effect.

At currents above the critical one, the laser emits just a few modes with the overall spectral width that cannot be attributed to ultrafast laser dynamics. The spectral behavior at currents exceeding the critical one is very different for the two measurements. (Below we demonstrate a poor reproducibility of the spectral behavior above the critical current.)

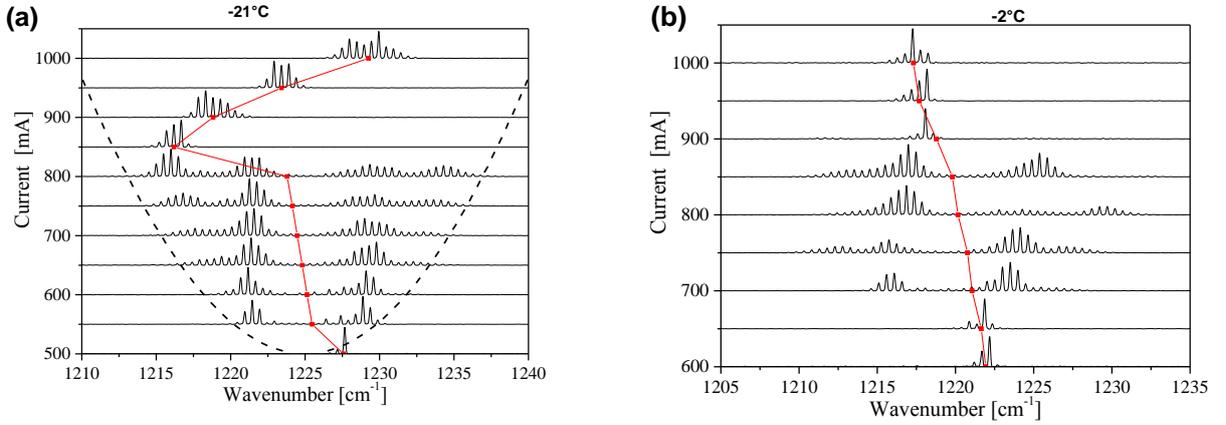

**Fig. 3.** Optical spectra of QCL sample #1803 at -21°C (a) and -2°C (b) heatsink temperature under cw operation regime. The red curves indicate the central wavelength. The dashed black curves in (a) indicate $\pm\Omega_{Rabi}/2\pi$ modulation sidebands calculated for $T_1$=0.7 ps, $T_2$=0.14 ps and D=180 cm$^2$/s.

The spectral behavior of the sample #6326 replicates the main features seen in the sample #1803, including the occurrence of the ultrafast multimode dynamics at small (~10 %) excess above lasing threshold and its collapse at a critical current of 0.8-0.85 A. Figure 4 shows the spectra measured at -25°C temperature while Figs. 7(a) and (b) [black curves] show these at the temperatures -17°C and once again at -25°C (repeated measurements from Fig. 4 to check the reproducibility). The apparent correlation of the critical currents in the two samples attests that the collapse of the ultrafast dynamic regime is due to the QCL multilayer structure design and composition.

In contrast to the good reproducibility of the multimode spectra in the RNGH self-pulsation regime and of the critical collapse current, we spot a poor reproducibility of the spectral behavior at currents above the critical one. For example in Figs. 4 and 7 (b), the very different spectra have been measured in function of the pump current in the sample #6326 at nominally the same heat sink temperatures but on different days. Such weak reproducibility of the spectra at a current above the critical one can also be a reason for the very different spectral behaviors observed above the critical current at various temperatures (compare Figs. 3, 4 and 7). We attribute it to an unstable operation with EFDs formation due to the misalignment of the otherwise resonant energy levels in the injector and active region [30,31]. Indeed, switching to EFD regime has a clear boundary which is mostly determined by the subband energy structure of the laser, and not by its temperature. Moreover, this mechanism allows us to explain the collapse of the multimode RNGH instability by a contrast reduction of the induced population inversion grating as a result of the carrier distribution and bias field inhomoheneities caused by EFDs formation.

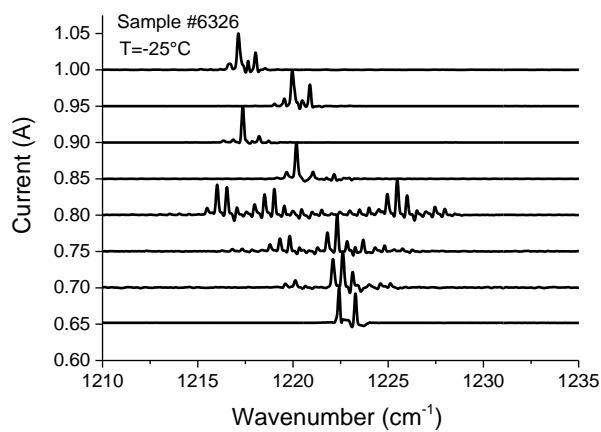

**Fig. 4.** Optical spectra of the sample #6326 at -25°C under cw pumping. [To be compared with Fig. 7(b)].

In order to support our interpretation of the observed QCL spectral behavior and its relation with RNGH instability, we study the second-order interferometric autocorrelation (IAC) traces as a function of the pump current. The IAC traces are measured using the step-scan mode of the FTIR spectrometer and a modified near-infrared InSb photovoltaic detector (D413, Bruker) with the sensitivity cut-off at 5.4 μm wavelength. We have replaced its sapphire window by a ZnSe one in order to use it in the two-photon absorption (TPA) measurements at 8 µm wavelength, as described in Ref [12]. In addition, an optical chopper with a low duty cycle, low noise current amplifier (SR570, Stanford Research Systems) and a lock-in amplifier (SR830, Stanford Research Systems) are used in the setup for suppression of spurious light background and detector heating as well as for increasing the signal-to-noise ratio.

Two representative IAC traces are shown in Figs. 5(a) and (b) (bottom panels). They are measured, respectively, at the currents just below the critical value and just above it. Figure 5 (a) corresponds to the case when a large spectral broadening of the laser emission is observed, while Fig. 5 (b) corresponds to the collapse of the ultrafast dynamic regime (the top panels in Fig. 5 show the corresponding spectra).

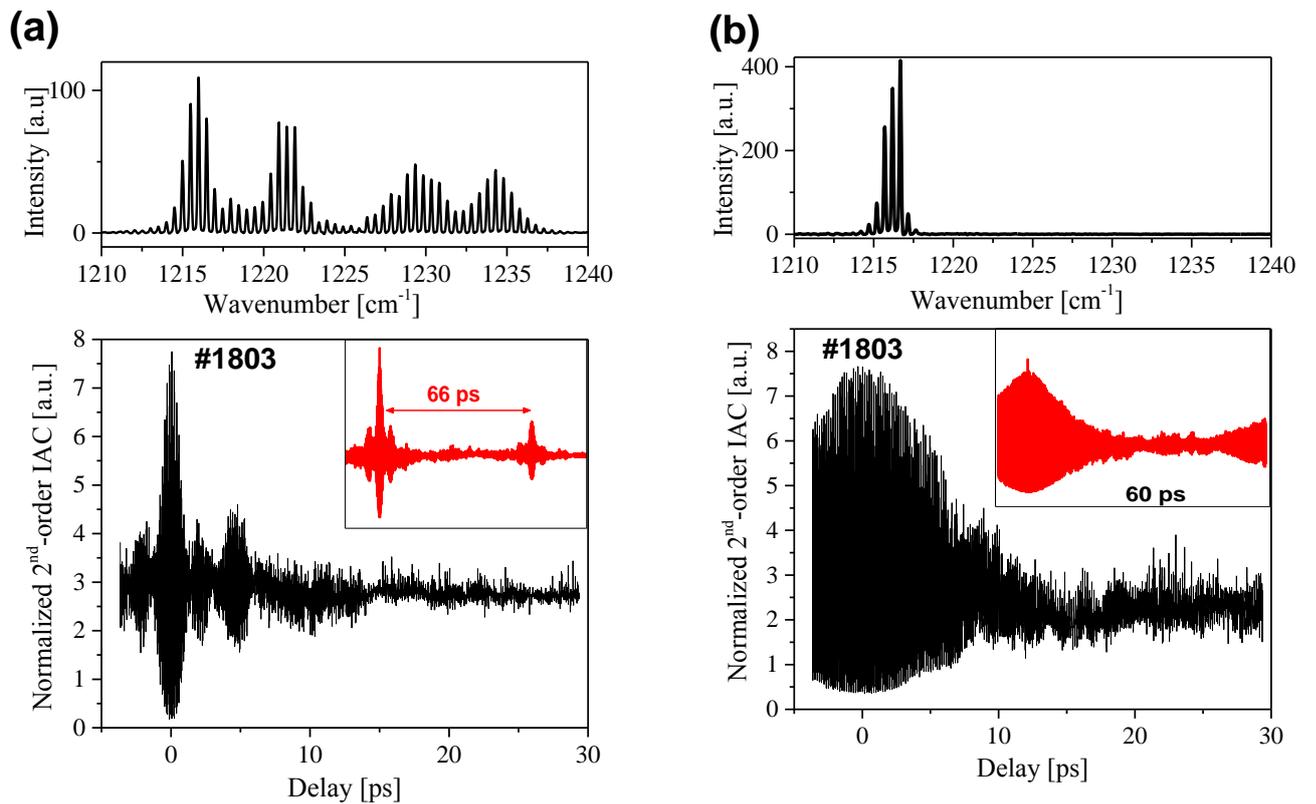

**Fig. 5.** Optical spectra (top panels) and second-order interferometric autocorrelation (IAC) traces (bottom panels) of the sample #1803 at just below (a) and just above (b) the critical current of the ultrafast multimode regime collapse. The inset shows the IAC trace for longer delay time. QCL heatsink temperature is -20°C.

At a current just below the collapse of the ultrafast multimode dynamics, the peak-to-background ratio of the main IAC lobe at zero delay is close to 8:3 [Fig. 5(a)]. The replica of the main lobe at a delay of one cavity roundtrip (66 ps) is of a largely reduced amplitude [see the inset in Fig. 5(a)] attesting that the relative phases of the modes do not completely recover after one cavity roundtrip. For a perfect mode-locking regime, the peak to background ratio would be of 8:1 [13], and the delayed lobe would be of the same amplitude as the main one. The IAC trace in Fig. 5(a) rather evidences for a "noisy CW" (stochastic) process [13]. For example, a quasiperiodic chaotic pulse train numerically modelled in Ref. [7] for a QCL with the cavity length of a few millimeters, can reproduce a similar IAC contrast. The model in [7] predicts an intermittent behavior of the optical wave field with several sporadic alternation of the field sign on the cavity roundtrip. In Fig. 5(a), the half-width at half-maximum (HWHM) of the central lobe measured above the background [at the level of 5.5 in Fig. 5(a)] is 0.6-0.7 ps. This duration corresponds to the transient processes associated with the field sign changes in the modelled optical waveform in Ref.[7].

The secondary side-lobes in the IAC trace in Fig. 5(a) at a delay of 4.7 ps, which is much smaller than the cavity roundtrip time, can be attributed to the multiple field sign alternations on a cavity roundtrip. The quasi-periodicity of such alternations is evidenced by the side-lobe structure which is recovered in the delayed replica at 66 ps. Similar features have been experimentally observed in Ref.[2] and are reproduced by the numerical model of Ref.[7].

At a current just above the collapse of the ultrafast multimode dynamics (Fig. 5(b)), the central IAC correlation lobe broadens without a noticeable change in the peak-to-background ratio of the central fringe. The multimode laser dynamics thus remains in the class of "noisy CW" processes [13]. However, this time no secondary side-lobe features can be seen within the cavity roundtrip indicating that the waveform does not have a quasi-periodic internal structure. The lasing dynamics becomes much simpler and slower and can be associated with emission of several uncorrelated cavity modes.

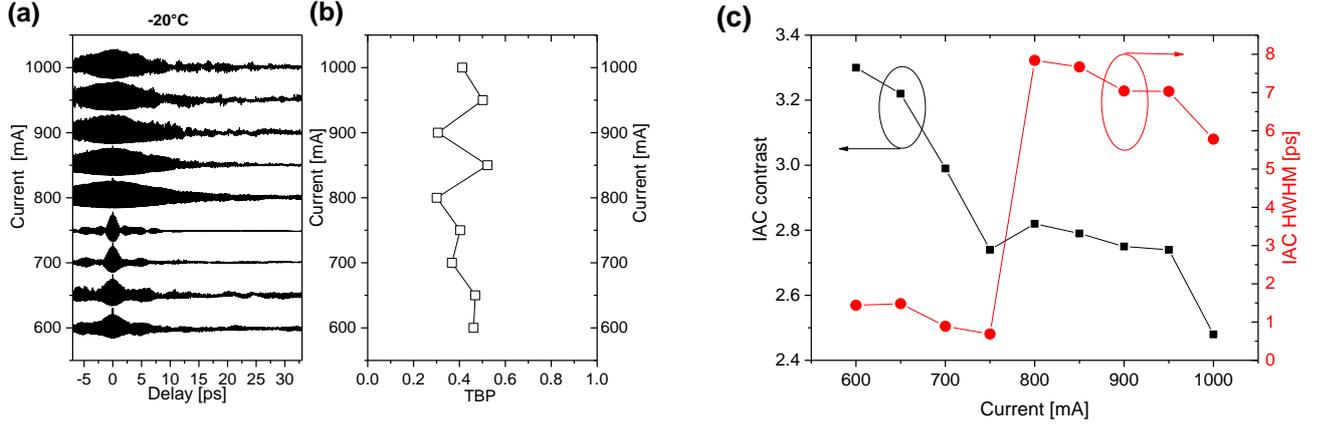

**Fig. 6.** Second-order Interferometric autocorrelation (IAC) traces of the sample #1803 (a), time-bandwidth product (b), IAC central lobe contrast [(c), left axis] and HWHM of the central IAC lobe measured above the background [(c) right axis] in sample #1803 as a function of the pump current at the heatsink temperature -20°C.

Figures 6 (a), (b) and (c) provide an insight into QCL behavior in the time domain as a function of the pump current, and show the variations of the second-order IAC patterns, time-bandwidth product (TBP), IAC peak-to-background ratio (IAC contrast) and the half-width at half maximum (HWHM) of the central IAC lobe. The bandwidth is estimated from the data in Fig. 3(a). Note that just above the lasing threshold (at about 500 mA) when the laser emits a very few modes [see Fig. 3(a)], the output power is insufficient to record the second-order IAC trace using our setup. The first data trace in Fig. 6 is taken at the pump current already above the second threshold (at about 550 mA) when the spectrum reveals a broadband multimode emission.

With increasing the current, the central IAC lobe [Figs. 6(a) and (c), right axis] narrows until the current reaches the critical value where the ultrafast multimode dynamics collapses. The narrowing of the central lobe and the reduction of the characteristic repetition time of the side-lobe structure attests for an acceleration of the intermittent transients in the optical waveform, in agreement with the increase of the overall spectrum width (Fig. 3).

The highest peak-to-background ratio of 8:2.4 (~3.3) in the IAC trace has been observed at the smallest current above the second threshold (at 600 mA) with a couple of secondary side-lobes attesting for the formation of a quasi-periodic waveform [7]. With increasing current, the peak-to-background contrast ratio reduces down to 8:3 (~2.7) that can be associated with increasing number of irregular field sign alternations at one cavity roundtrip. Above the critical current, the peak-to-background ratio remains near 8:3, which is typical for a noisy CW lasing regime [13] with uncorrelated multimode emission.

While the shape of IAC, its contrast, and the HWHM width (Fig. 6(c), right axis) clearly show a change in the dynamic regime at the critical current (750-800 mA), the TBP does not indicate any signature of an abrupt transition. The TBP always remains low, around a value of 0.4-0.5, indicating that the waveform (the pulse shape) cannot be improved with a dispersion correction. These is because the waveform remains irregular at any current, whether it is a quasi-periodic chaotic pulse train or a noisy multimode CW emission. This conclusion is consistent with the fact that none of free-running QCL frequency combs has been transformed into a distinct pulse train in the time domain.

Thus, the IAC measurements confirm our assumption that the multimode dynamics responsible for the strong spectral broadening with the width of the order of Rabi oscillation frequency is due to excitation of ultrafast quasiperiodic RNGH self-pulsations predicted by the numerical model of Ref. [7]. We now will provide an additional consideration supporting the picture of unstable EFDs formation at currents above the critical one.

**B. Effect of optical pumping in steady state regime**

In order to confirm the proposed scenario of RNGH instability collapse, we compare the optical spectra in free-running QCL #6326 (black spectral traces in Fig. 7) and in the same QCL subjected to an additional optical pumping from a mode-locked Ti:S laser tuned to 860 nm wavelength (red spectral traces in Fig. 7). For fidelity, the measurements in Fig. 7 are repeated at two different temperatures [compare the panels (a) and (b)].

The Ti:S laser pulses of 80 fs duration with the repetition rate 80 MHz are stretched in a single-mode optical fiber to 11 ps width and focused at the output facet of the QCL. A thin Si plate is used as a dichroic beam splitter to combine/separate the QCL output and the pump beams. The average power of the excitation beam measured before the QCL collimation lens does not exceed 100 mW. The pump photon energy corresponds to the conduction-valence band transition energy above the QW barriers. The photons are absorbed in a narrow region nearby the output facet of QCL. Such optical pumping on conduction-valence band transition allows us to generate locally free non-equilibrium carriers.

The interband pumping generates electron hole pairs with the energy distribution dependent on the pump photon energy. A significant influence of the inter-band optical pumping on the output power near the lasing threshold in InGaAs/InAlAs on InP QCLs have been observed [17]. The interband pumping with the photon energies matching the lasing subband edge in the active QWs (~1.6-1.7 µm wavelength) populates the lasing subbands directly and enhances the output emission power. However, the optical pumping at the wavelength tuned to the bandgap in the QW barriers (about 850 nm) creates hot electrons and holes in highly energetic states and causes a reduction of the output power. Such reduction of emission power has been attributed in [17] to depopulation of the upper lasing level by scattering with hot electrons. However, one important factor was omitted in these considerations. Namely the generated non-equilibrium holes in *n*-type InGaAs, InAlAs, and InP have very long non-radiative lifetimes (of the order of few µs). These lifetimes are much longer than the pump pulse train period of 12.5 ns. As a result, the generated locally free holes are accumulated in the QCL sample, and they are distributed inhomogeneously along the waveguide because of the edge-pumping configuration and due to a final diffusion length (on the order of 100 µm). The non-equilibrium holes produce screening of the applied bias field and increase the inhomogeneity of the bias field along the waveguide. In addition, the non-equilibrium holes may result in free-carrier absorption.

Our experiments in Fig. 7 thus show that inter-band optical excitation of InGaAs/InAlAs on InP QCLs in the edge-pumping configuration and with the photon energies above the QW barriers results in (i) a small reduction of the overall output power at any pump current (like the power drop observed in [17]) and in (ii) the collapse of RNGH instability at a slightly smaller critical current than in free-running QCL. No significant change in the character of the optical spectra is observed at other currents. Both effects can be attributed to an inhomogeneity of the internal field and of the gain and loss distributions along the cavity caused by the locally excited non-equilibrium holes. At a current near critical, this inhomogeneity adds up to that of EFDs, reducing the contrast of SHB-induced grating and facilitating the collapse of RNGH self-pulsations.

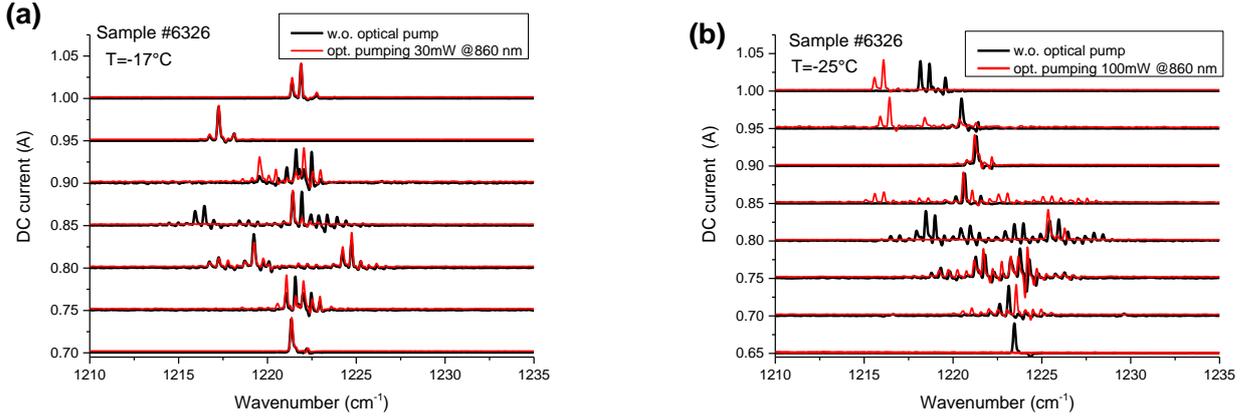

**Fig. 7.** Optical spectra of the sample #6326 at –17°C (a) and -25°C (b) under cw pumping with and without optical pumping by mode-locked Ti:S laser tuned at 860 nm. The average power incident on the laser facet measured before the collimation lens was 30 mW(a) and 100 mW (b).

**C. Transient turning-on behavior**

Now we turn to the switching between the diagonal and vertical transitions having different relaxation times $T_1$. This switching occurs in the beginning of the pump current pulse and can lead to the changeover of the multimode RNGH instability. We measure the transient behavior in the QCL sample #6326 driven by 10 µs current pulses with the rising front duration <0.1 µs. The current pulses are obtained by discharging a 10 µF capacitor via a series resistor of 100 Ω (the characteristic discharge time is 1 ms). The circuit is commutated to the QCL via a high-power MOSFET switch (IRFPG50, Vishay Siliconix) with the switching time of 35 ns. (For comparison, the drain-source resistance of the turned-on MOSEFT switch is 2 Ω and the differential resistance of the QCL above the threshold is ~15 Ω). In order to produce square 10 µs current pulses, the MOSFET switch is controlled by a waveform generator (WW2571A, Tabor). The current amplitude is set by the initial charge of the capacitor.

The time-resolved spectra are recorded using the step scan mode of the FTIR spectrometer (5 ns data sampling rate) and a linear-response MCT detector (D316, Bruker). The temporal resolution is limited by the detector bandwidth of 20 MHz (17.5 ns response time).

Figures 8 and 9 show the time-resolved spectra taken at the heatsink temperatures of -18°C and 0°C, respectively, in QCL operating just above the lasing threshold [Fig. 8(a) and (b)], in the regime of RNGH self-pulsations [Figs. 8(c) and 8(d), 9(a)], and above the critical current of the RNGH instability collapse [Figs. 8(e) and (f), 9(b)]. The quoted current amplitudes in Figs. 8 and 9 for specific dynamic regimes differ from those in Figs. 4 and 7 because of the transient behavior of the carrier distributions. In Figure 8 we also show the integrated intensity profile and the behavior of the peak emission wavelength (highest intensity mode) extracted from the time resolved spectra.

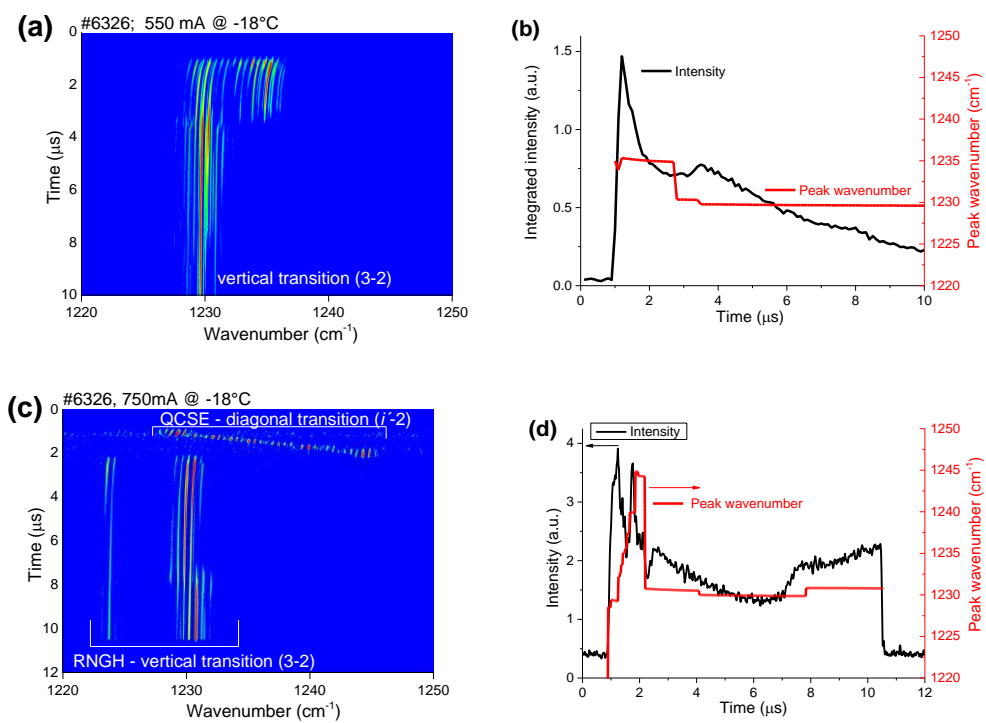

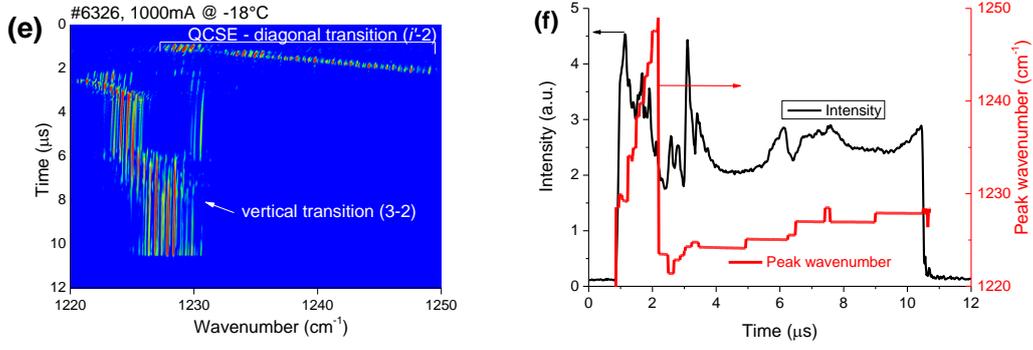

**Fig. 8.** Spectrochronograms (left panels) and extracted integrated intensities and peak emission wavelengths (right panels) in the free-running QCL sample #6326 measured under pumping with 10 µs current pulses of the amplitude 550 mA (a) and (b), 750 mA (c) and (d) and 1000 mA (e) and (f) at the heatsink temperature of -18°C. The spectral and temporal resolutions are 0.5 cm$^{-1}$ and 20 ns, respectively.

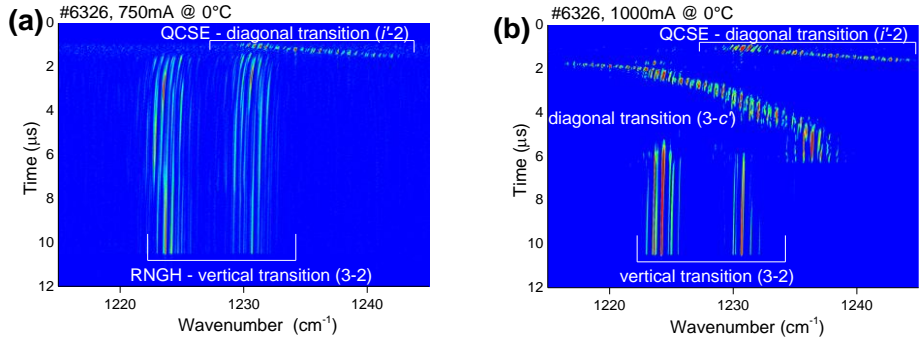

**Fig. 9.** Spectrochronogram of the free-running QCL sample #6326 measured under pumping with 10 µs current pulses with 750 mA (a) and 1000 mA (b) amplitude and the heat sink temperature of 0°C. The spectral and the temporal resolutions are 0.5 cm$^{-1}$ and 20 ns respectively.

In Figs. 8 and 9, the current pulse is applied at 1 µs time on the time axis. We distinguish three phases in the output optical pulse. An accumulation of non-equilibrium carriers and the onset of lasing emission occurs in the initial ~0.1 µs interval. Its exact duration is defined by the rising front of the emission pulse until the moment when the integrated intensity reaches a maximum.

Within the next ~1 µs interval we observe a narrow-band lasing with the blue-shifted and up-chirped emission spectrum and numerous mode hops. During this second phase, the integrated intensity decreases roughly by a factor of two.

In the last portion of the pulse and for the most of driving conditions, the laser switches to another transition. The laser reveals a multimode regime with either a broadband spectrum consisting of the two mode clusters [Figs. 8(c), 9(a) and (b)] or just an emission with several uncorrelated modes in the optical spectrum [Fig. 8(a) and (e)]. The individual modes show a slow red shift due to self-heating.

Two deviations from this picture have been noticed. Firstly, at a current just above the lasing threshold and below the second threshold [Fig. 8(a)], no up-chirped lasing emission occurs on a separate transition in the beginning of the current pulse. Secondly, at a high heatsink temperature and high pump current [Fig. 9(b)], prior to the broadband multimode emission with two clusters of modes, the lasing occurs on another yet blue-shifted and up-chirped transition.

A somewhat similar behavior of the emission wavelength under a pulsed pumping with the rapid up-chirp of the frequency in a few microseconds in the beginning of the pulse, and its subsequent rapid drop followed by a smooth glissade due to self-heating effects was observed in THz QCLs [18, 19]. The up-chirp in the beginning of the applied current pulse was attributed in [18] to a reduction of the population inversion on the lasing transition yielding a change of the imaginary part of the dielectric constant. However, the laser in [18] was not a single-mode laser and hence the refractive index change cannot explain the behavior of the envelope of the entire emission spectrum.

Interestingly, in Figs. 8, within the first microsecond of the pump pulse, we also observe a rapid decrease of the overall output power that accompanies up-chirped emission. Such output power behavior does provide an evidence of the reduction of the pump rate on the lasing transition.

According to the transition energy diagram in Fig. 2(b), the multimode emission in the last (near steady-state) portion of the pulse occurs on the main vertical transition 3-2 [black curve in Fig. 2(b)], while the up-chirped blue-shifted emission on the initial portion of the pump pulse can be attributed to the diagonal transition $i'$-2 [red curve in Fig. 2(b) in the range of bias fields 32-34 kV/cm]. The lasing observed in Fig.9(b) with the photon energies between these two lines can be attributed to the emission on another yet diagonal transition 3-

$c'$ [green curve in Fig. 2(b)]. Note that the QCSE energy shift of the 3-$c'$ line is weaker than that of the $i'$-2 line, which allows us to distinguish two diagonal transitions in the experiment. Small discrepancies between the experimentally observed spectral behavior of different lines in Fig. 9(b) and the one predicted by the model in Fig.2 (b) can be attributed to that fact that the model assumes a steady state carrier density, the while during the power turning-on experiment, QCL undergoes a transient process with a non-stationary carrier distribution.

We therefore attribute the observed up-chirp of the lasing emission on the diagonal transitions to an increasing bias field within the first microsecond of the pump pulse. In order to affirm this attribution, we shall demonstrate that the decrease of the pump rate on the lasing transition occurs simultaneously with the increase of the bias field in our experiment. Unfortunately the directly measured current and voltage waveforms exhibit strong ringing behavior at 40 MHz frequency within the first microsecond of the current pulse. At the same time no any ringing trace is observed in QCL emission in Figs. 8(b), (d) and (f), indicating that it occurs due to parasitic capacitance and inductance of QCL package and contact wires. Nevertheless it is quite strong and completely masks the actual current and voltage variations across the QCL active region.

Fortunately a set of modelled I-V curves like the one in Fig. 2(c) allows us to qualitatively explain the observed spectral behavior in the turning–on QCL. Using that the photon induced carrier transport [16] has a minor effect on the I-V curve at a room temperature, we notice that the I-V curve in Fig. 2(c) is obtained for a steady-state equilibrium distribution of the carriers and a fixed temperature. At the same time, in the beginning of the applied current pulse, a large population inversion is created on the main vertical transition and on the diagonal transitions. During the first microsecond of the current pulse, QCL operates with carrier densities significantly higher than that in the steady-state operation regime. This provides us a clue to the observed spectral behavior.

In Figure 10 (a) we plot the I-V curves (in terms of the total current density in function of the bias field) at the equilibrium carrier density defined by the doping level [the same as the blue curve in Fig.2(c)], as well as at 2 and 3 times higher overall carrier densities. The I-V curves are shown together with the load line of the current source (dashed green line). The initial operation point with high carrier density is indicated with the label "*I*". The operation point "*F*" for a steady-state carrier density is located on the load line at a smaller current and, respectively, at a higher voltage bias. The observed QCL spectral behavior in the first microsecond of the current pulse can be explained by the non-equilibrium carrier density that relaxes towards the steady state [from "*I*" to "*F*" curve in Fig. 10 (a)] along the load line of the current source. As seen from Fig. 10 (a), this causes a decrease of the overall current (and of the pump rate) and an increase of the bias field. According to Fig.2 (b) that should produce a strong up-chirp in the diagonal transition $i´$-2 (and 3-$c´$), which indeed has been observed in the experiment (see Figs. 8 and 9) together with the reduction of the output power.

In Fig. 10 (b) we plot the total current normalized to the threshold currents for a few transitions of interest. We consider both the steady-state operation regime "*F*" (closed symbols) and the initial operation with non-equilibrium carrier densities "*I*" (open symbols). Note that the threshold current changes with the bias field because the injector-active region level alignment varies, producing a change in the matrix elements, lifetimes and population inversions. When the total current exceeds the lasing threshold for a particular transition (the corresponding curve in Fig. 10 (b) passes above the value of 1 indicated by the dotted horizontal line), QCL may potentially lase at this transition. For example, QCL can lase on the main vertical transition 3-2 at equilibrium carrier density starting from the bias field of ~ 32 kV/cm. When the carrier density is increased (as in the begging of the pump pulse), the lasing on the diagonal transition $i´$-2 (and 3-$c´$) becomes possible at the bias field around ~ 32 kV/cm that is near the steady-state lasing threshold for the vertical transition 3-2. At the same time in the steady-state operation, these diagonal transitions exhibit insufficient small-signal optical gain for lasing. These model predictions are in agreement with experimental observations. Figure 10 (b) shows also that in a steady state, lasing on the diagonal transition $i$-2 can be possible in the range of the bias field 33-34 kV/cm. However these considerations do not take into account the population clamping effect in a steady-state operating laser. In our case it prevents lasing on the diagonal transition $i$-2 once the steady-state emission on the vertical transition 3-2 has been reached. So far our model has allowed us to identify and explain the observed set of the lasing transitions and relate these to particular range of the bias fields. However we should also acknowledge that the model does not explain yet why the lasing starts on the diagonal transition and not on the vertical one. We attribute this to the fact that the I-V curves in Fig. 10 are obtained assuming non-equilibrium but stationary carrier populations, while in the experiment, the system undergoes a transient process.

In Fig. 10 (c) we analyze the excess of the total current above the 2$^{nd}$ threshold (multimode RNGH instability threshold) for each of the selected transitions. The diagonal transitions $i´$-2 and 3-$c´$ which lase in the first microsecond of the current pulse at a high non-equilibrium carrier density and ~32 kV/cm bias field do not reveal RNGH instability. Their 2$^{nd}$ threshold currents are basically higher than the total pump current. Once the initial carrier density relaxes to the equilibrium one, the broadband multimode emission due to RNGH instability can occur in the steady-state operating QCL in Fig.10(c) only when QCL is lasing on the main vertical transition 3-2, exactly in the way as this was observed in experiment. (As discussed above, the transition $i$-2 shall be excluded in virtue of the population clamping).

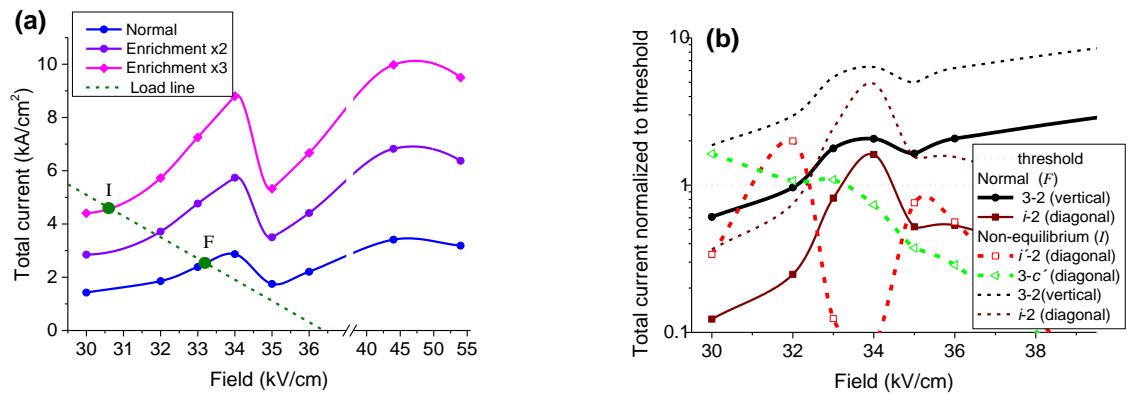

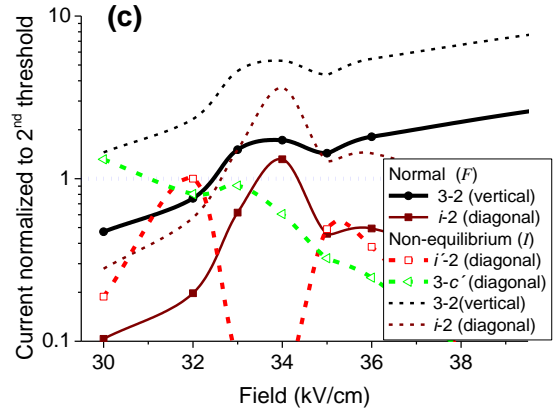

**Fig. 10.** (a) Effect of the carrier density on I-V curve of QCL. (b) Total current normalized at lasing threshold currents for the vertical 3-2 and a few diagonal transitions for a steady state (normal) carrier density as well as for the initial non-equilibrium carrier density. Only transitions capable of lasing are indicated. (c) Total current normalized to the 2$^{nd}$ threshold for the same transitions as in (b)

## V. Conclusion

In summary we have shown that FP cavity QCL with carrier transport by sequential resonant tunneling exhibit very reach multimode dynamics. The onset of broadband multimode emission associated with RNGH instability in QCLs can be tailored with applied bias field, via transient non-equilibrium and non-homogeneous distributions of carriers as well as via switching between the lasing transitions. We have explained all experimentally observed changes of the dynamic regimes in QCL using a simple analytical expression for the second threshold . The

Acknowledgement:

This research was supported by the Swiss National Science Foundation (SNF) project FASTIQ (ref. no. IZ73Z0_152761) as well as the European Union's Horizon 2020 research and innovation programme under the grant agreement No 686731 (SUPERTWIN), COST action BM1205 and by the Canton of Neuchâtel. QCL samples used in this research have been provided by Alpes Lasers.